\documentclass[]{aa}  

\usepackage{graphicx}
\usepackage{xcolor} 
\usepackage{txfonts}
\usepackage{lipsum}
\usepackage{subcaption}         
\usepackage{lscape}             
\usepackage{placeins}           
\usepackage[colorlinks=True]{hyperref}

\hypersetup{
    colorlinks=true,
    linkcolor=blue,
    filecolor=blue,      
    urlcolor=blue,
    citecolor=blue
}

\newcommand{\oiii}{[\ion{O}{III}]\,}
\defcitealias{Bicknell+97}{B97}
                                
\begin{document}

   \title{Radio Monitoring of the Changing-look AGN Mrk 590 I: VLA Observations Reveal A Bow-shock Driven Radio Brightening Event}


   \author{Gregory Walsh\inst{1}\fnmsep\thanks{\email{gregory.walsh@nbi.ku.dk}}
        \and Marianne Vestergaard\inst{1,2}
        \and Daniel Lawther\inst{1}
        \and Sandra I. Raimundo\inst{3,1} 
        \and Jun Yi Koay\inst{4,5} 
        }

   \institute{DARK, Niels Bohr Institute, University of Copenhagen, Jagtvej 155, 2200 Copenhagen N, Denmark
            \and Steward Observatory and Dept. of Astronomy, University of Arizona, 933 N. Cherry Avenue, 85721 Tucson, AZ, USA
            \and Physics and Astronomy, University of Southampton, Highfield, Southampton SO17 1BJ, UK
            \and Curtin University, Kent Street, Bentley, Western Australia 6102, Australia
            \and Institute of Astronomy and Astrophysics, Academia Sinica, 11F of Astronomy-Mathematics Building, AS/NTU No. 1, Sec. 4, Roosevelt Road, Taipei 106216, Taiwan, R.O.C.
            }


 
  \abstract
   {Mrk 590 is a changing-look AGN (CLAGN) that slowly faded from a Seyfert Type 1 to near quiescence over a few decades before re-igniting in 2017. Monitoring in the optical/UV/X-ray wavebands has revealed accretion-driven variability signatures, but the response of the radio source to the changing-look events that drove its fading and re-ignition is unknown. We present multi-frequency (1-17~GHz), A-configuration, Karl G. Jansky Very Large Array (VLA) observations of Mrk 590 over four epochs from 2015 to 2017. The radio source is dominated by a faint, compact component, absent of any notable extended structures. We identify a variable radio spectrum over this period, both in amplitude and in shape, and implement a Bayesian inference modeling routine to constrain the mechanism driving the radio variability. Our routine favors radio variability powered by a bow shock produced through an interaction of a radio jet with the interstellar medium, which drives the radio brightening in later epochs. This demonstrates that the radio variability we observed is not correlated with the variable accretion event that led to the re-brightening of the AGN at higher energy wavebands. While correlated variability between the radio and accretion-driven emission has been observed in other CLAGN, the lack of correlated variability in Mrk 590 shows the diversity of changing-look AGN populations and the mechanisms driving their behavior.}


\authorrunning{G. Walsh et al.}
\titlerunning{A Shock Front Drives Radio Variability in Mrk 590}
   \maketitle
   \nolinenumbers

\section{Introduction}
Contemporary time-domain, all-sky surveys at wavebands across the electromagnetic spectrum have identified a new sub-class of active galactic nuclei (AGN). These AGN exhibit extraordinary high amplitude photometric variability, characteristically in the optical, UV and/or X-ray bands, over time scales of months to years \citep[][]{Osterbrock_81,Penston&Perez_84,Stochi-Bergmann+93,Eracleous&Halpern_03,Risaliti+07,McElroy+16,Graham+20}. The photometric variability of these so-called changing-look AGN (CLAGN) is often coherent across the observed wavebands and drastically exceeds the variability amplitude associated with the stochastic variability that characterizes typical AGN. In some cases, the coherent photometric variability is accompanied by a change in the strength and profile of the broad emission lines. The (dis)appearance of the broad lines marks a transition between Seyfert type classifications which, for a number of these AGN, has been shown to be a direct result of a change in the underlying accretion flow itself \citep[][]{Denney+14,McElroy+16,Husemann+16,Trakhtenbrot+19,Ricci+20,Lawther+25} rather than a change in the line-of-sight obscuration. These changing-state AGN (CSAGN) have altered our understanding of the physics of the accretion onto supermassive black holes.

While much work has been dedicated to searching for and examining the correlated variability of the optical/UV/X-ray emission \citep[][]{Rumbaugh+18,Trakhtenbrot+19b,Timlin+20,Shen_21,Lawther+23,Veronese+24,Lawther+25}, which directly probes the accretion disk and corona of the AGN \citep[][present a recent review]{R&T_23}, the radio properties of the CLAGN population still remain mostly unexplored. \citet[][]{Birmingham+25} examined the radio light curves of a sample of 474 spectroscopically-confirmed CLAGN using the Australia SKA Pathfinder Variable and Slow Transients Survey, and the Very Large Array Sky Survey. They did not identify systematic fading or enhancement of the radio flux density after a changing-look event at the population-wide level. Additionally, the occurrence of low radio amplitude variability is equally likely between the CLAGN and non-CLAGN control population in their sample.
The lack of correlated variability between the radio and optical/UV/X-ray sources suggests that the mechanisms responsible for the variable accretion rate do not significantly alter the radio emission mechanism. Alternatively, the time scale over which the accretion-driven radio variability occurs could be much longer than what is found for the accretion-driven variability at higher energies.

Dedicated programs targeting individual CLAGN have shown tantalizing evidence to suggest a causal connection between the radio source and accretion-driven emission. Most notably, the CLAGN 1ES 1927+654, which lost and reformed its X-ray corona over a few months \citep[][]{Trakhtenbrot+19, Ricci+20}, has exhibited a 60-fold increase in its multi-frequency radio flux density beginning approximately 1800 days following the changing-look event, as revealed by Very Long Baseline Array (VLBA) observations \citep[][]{Meyer+25}. 1ES 1927+654 has sustained this level of enhanced flux density for over a year following its exponential rise. The brightening of the VLBA source has been accompanied by both the appearance of a convex radio spectrum and the launching of a mildly relativistic bipolar outflow \citep[][]{Meyer+25}. X-ray spectral and timing studies have suggested that the changing-look event observed for 1ES 1927+654 was powered by a tidal disruption event (TDE) in a pre-existing AGN accretion disk \citep[][]{Masterson+22,Masterson+25}. The larger population of TDEs have been observed to exhibit radio emission brightening on time scales longer than a few hundred days after their discovery \citep[][]{Cendes+24,Alexander+26}, suggesting that the late-time radio response of this changing-look event is analogous to what is observed in off-axis jetted TDEs.

Some CLAGN have shown multiple accretion state transitions over decades of monitoring. Mrk 1018 has transitioned from a Seyfert Type 1.9 \citep[][]{Osterbrock_81} to a Seyfert Type 1 \citep[][]{Cohen+86} and back to a Seyfert Type 1.9 \citep[][]{McElroy+16} over the past 40 years \citep[][]{Dunn+25}.
Optical, UV, and X-ray studies have identified systematic variations at these wavebands that accompany the transitions \citep[e.\,g,][]{Lyu+21,Brogan+23,Veronese+24}. The overall dimming of the source as it transitioned back into a low accretion state is consistent with the innermost disk evaporating and forming an advection dominated accretion flow \citep[][]{Noda&Done_18}. Around this time, \citet[][]{Walsh+23} identified proper motion of the VLBA radio source that is consistent with the ejection of a new outflowing component to the radio jet. Although this may suggest correlated variability properties of the radio and accretion-driven emission, the lack of long-term, coordinated monitoring of Mrk 1018 limits the capabilities to establish a causal connection between the radio source and the accretion flow.

For the analyses presented here, we focus on the CLAGN \object{Mrk 590} ($z=0.026385$). This AGN was a typical Seyfert Type 1 before the year 2012, at which point the optical AGN continuum and broad emission lines faded \citep[][]{Denney+14}. Although low in X-ray flux by 2012, the soft X-ray excess was not completely absent during Mrk 590's dim state \citep[][]{Denney+14,Mathur+18}. This suggests that the AGN was already refueling by 2014, and, indeed, its optical broad emission lines reappeared in 2017 \citep[][]{Raimundo+19}. \citet[][]{Lawther+23} reported repeated UV/X-ray flaring events with a characteristic time scale of $\sim100$~rest-frame days beginning in 2017, the origin of which remains unknown. \citet[][]{Koay+16b} examined the long-term radio light curve of Mrk 590 between the years 1970-2015 and found a systematic dimming of the radio flux densities as the AGN transitioned to a dim flux state. However, no such comprehensive analysis of Mrk 590's near-term radio variability, i.\,e., following its reawakening in 2017, has been completed.

In this paper, we examine the radio variability of the CLAGN Mrk 590 through observations conducted with the Karl G. Jansky Very Large Array (VLA) between 2015 and 2017 (Section~\ref{sec:observations}). We present the radio source characteristics identified through both stacked imaging (Section~\ref{sec:source_params}) and time-series analyses (Section~\ref{sec:modeling}). We perform Bayesian inference modeling of the time-series radio spectra (Section~\ref{sec:modeling}) and assess the possible mechanisms driving the radio variability in 2016 and 2017 (Section~\ref{sec:var_origins}). Lastly, we discuss the association of the radio variability with the changing-look event (Section~\ref{sec:corr_var}) and present our conclusions (Section~\ref{sec:conclusions}).

\section{Observations and Data Reduction}\label{sec:observations}
Motivated by the changing-look behavior of Mrk 590, first reported by \citet{Denney+14}, we obtained observations of Mrk 590 with the VLA to assess if the radio source exhibited correlated variability with that at the optical-UV-X-ray wavebands following the AGN's decline from a Type I state. 
Our VLA observations of Mrk 590 were carried out over two separate A-configuration programs: a single epoch on 2015 June 23 and three epochs between 2016 October 5 to 2017 January 10 under the programs 15A-084 (PI: J.Y. Koay) and 16B-382 (PI: J.Y. Koay), respectively. Consistency between the synthesized beam sizes of the two programs ensures us that any variability of the source flux density is not a result of source confusion but is a genuine, astrophysical effect.
Both programs observed at central frequencies of 1.52~GHz (L-band), 5.75~GHz (C-band), 9.30~GHz (X-band) and 16.0~GHz (Ku-band). The 2015 observations also incorporated observations at 380~MHz (P-band), however we did not calibrate this data set or include it for our analysis as the low frequency of these observations is dominated by star-formation processes (see Section~\ref{sec:source_params}). 380~MHz observations were not included in the later three epochs of our program because of this and thus we exclude these single-epoch data from the current work focused on the nuclear radio emission properties. The receiving bands, their total bandwidth and central frequency are listed in Table~\ref{tab:vla_seimage_params}. Each frequency was observed in full polarization mode. Our observations used 3C 48 for flux density and bandpass calibration, and J0215$-$0222, located $1.64\deg$ away, for complex gain calibration. Our observations used 3C 84 for a reference pointing scan to correct for antenna pointing offsets at the beginning of each observation and again before 16.0~GHz scans of Mrk 590.

We calibrate the data using the Common Astronomy Software Applications package (\verb|CASA|; \citealt{CASA_22}) pipeline version 6.6.6-17. This routine automates the standard calibration procedures including flagging, Hanning smoothing, and flux density, bandpass and complex gain calibration. We inspect the calibrated visibility data of Mrk 590 and remove any remaining spurious data affected by radio frequency interference via automated and manual flagging routines. 

\begin{table*}[t!]
\centering
\caption{Mrk 590 VLA Single-Epoch Observation and Source Parameters}
\begin{tabular}{ccccccccc}
\hline \hline
\noalign{\smallskip}
Program ID & Date & Band & $\nu_\mathrm{cent}$ & $\Delta\nu$ & $S_\nu^\mathrm{peak}$ & $S_\nu^\mathrm{int}$ & Image RMS & $\theta_\mathrm{maj} \times \theta_\mathrm{min}$ \\
 & & & (GHz) & (GHz) & (mJy beam$^{-1}$) & (mJy) & (mJy beam$^{-1}$) & (\arcsec) \\
 (1) & (2) & (3) & (4) & (5) & (6) & (7) & (8) & (9) \\
\noalign{\smallskip}
\hline \\
15A-084 & 2015 June 23 & L & 1.52 & 1.024 & 3.20 $\pm$ 0.10 & 4.30 $\pm$ 0.10 & 0.037 & 1.83$\times$1.38 \\
-- & -- & C & 5.75 & 2.048 & 2.71 $\pm$ 0.08 & 2.76 $\pm$ 0.08 & 0.016 & 0.47$\times$0.34 \\
-- & -- & X & 9.30 & 2.048 & 2.96 $\pm$ 0.09 & 3.08 $\pm$ 0.09 & 0.012 & 0.30$\times$0.22 \\
\smallskip
-- & -- & Ku & 16.0 & 2.048 & 2.48 $\pm$ 0.14 & 2.55 $\pm$ 0.14 & 0.011 & 0.18$\times$0.13 \\

16B-382 & 2016 October 05 & L & 1.52 & 1.024 & 3.28 $\pm$ 0.11 & 5.39 $\pm$ 0.11 & 0.031 & 1.83$\times$1.24 \\
-- & -- & C & 5.75 & 2.048 & 3.65 $\pm$ 0.11 & 3.72 $\pm$ 0.11 & 0.016 & 0.51$\times$0.38 \\
-- & -- & X & 9.30 & 2.048 & 4.21 $\pm$ 0.13 & 4.27 $\pm$ 0.13 & 0.019 & 0.33$\times$0.24 \\
\smallskip
-- & -- & Ku & 16.0 & 2.048 & 4.32$\pm$ 0.13 & 4.37 $\pm$ 0.13 & 0.015 & 0.17$\times$0.12 \\

16B-382 & 2016 November 19 & L & 1.52 & 1.024 & 2.89 $\pm$ 0.10 & 5.23 $\pm$ 0.10 & 0.031 & 1.75$\times$1.18 \\
-- & -- & C & 5.75 & 2.048 & 3.19 $\pm$ 0.10 & 3.23 $\pm$ 0.10 & 0.015 & 0.45$\times$0.32 \\
-- & -- & X & 9.30 & 2.048 & 3.97 $\pm$ 0.12 & 4.03 $\pm$ 0.12 & 0.017 & 0.30$\times$0.21 \\
\smallskip
-- & -- & Ku & 16.0 & 2.048 & 4.86 $\pm$ 0.15 & 4.94 $\pm$ 0.15 & 0.014 & 0.16$\times$0.13 \\

16B-382 & 2017 January 10 & L & 1.52 & 1.024 & 3.34 $\pm$ 0.11 & 5.16 $\pm$ 0.11 & 0.031 & 1.78$\times$1.21 \\
-- & -- & C & 5.75 & 2.048 & 3.63 $\pm$ 0.11 & 3.70 $\pm$ 0.11 & 0.016 & 0.46$\times$0.35 \\
-- & -- & X & 9.30 & 2.048 & 4.22 $\pm$ 0.13 & 4.26 $\pm$ 0.13 & 0.019 & 0.30$\times$0.21 \\
-- & -- & Ku & 16.0& 2.048 & 4.32 $\pm$ 0.13 & 4.43 $\pm$ 0.13 & 0.013 & 0.16$\times$0.12 \\
\noalign{\smallskip}
\hline
\end{tabular}
\tablefoot{Program IDs (column 1), their observing date (2), frequency bands (3) from our observations and their central frequency (4) and bandwidth (5) for our VLA programs of Mrk 590 used in this paper. For each frequency band, we also report the single-epoch peak (6) and integrated flux densities (7) of Mrk 590, and the image RMS (8) and the restoring beam (9). The reported values for 1.52~GHz correspond to the central, unresolved component of the radio source.}
\label{tab:vla_seimage_params}
\end{table*}

After calibration, we split each band into two separate frequency ranges, divided into 8 spectral windows each, to increase the density of sampled data points in the radio spectrum of Mrk 590. We image the visibilities using the \verb|CASA| task \verb|tclean| with natural weighting, ensuring sensitivity to extended structure, and a Multi-Term Multi-Frequency Synthesis deconvolution algorithm \citep{Rau&Cornwell_11} to account for the large fractional bandwidths, exceeding 67\% for 1.5~GHz observations. This also provides us with a robust estimate for the in-band spectral index $\alpha_\mathrm{IB}$. The Stokes \textit{I} single-epoch images suffer from minor deconvolution artifacts due to the limited \textit{uv} coverage afforded by these snapshot observations, with each frequency attaining an on-source integration time of less than 10~minutes per epoch. For the 1.52~GHz and 16.0~GHz \textit{uv} data, we employ phase and amplitude self-calibration to improve the image quality, which increases the dynamic range of each image and sensitivity to low surface brightness features, if present. For 5.75~GHz and 9.30~GHz \textit{uv} data, phase-only self-calibration attained model convergence and the maximal image quality. Our self-calibration model also includes strong, nearby field sources that dominate at lower frequency ($<5$ GHz), leading to an improved dynamic range for each image. To maximize the sensitivity to intrinsically weak features, we stack the data sets in the \textit{uv} plane and produce a single image for all frequencies except for 16.0~GHz. Due to source variability (Section~\ref{sec:modeling}), our stacked image for the 16.0~GHz data sets features only the three epochs from the 16B program.

\section{Source Parameters}\label{sec:source_params}
\begin{figure*}[t!]
    \centering
    \includegraphics[width=\columnwidth]{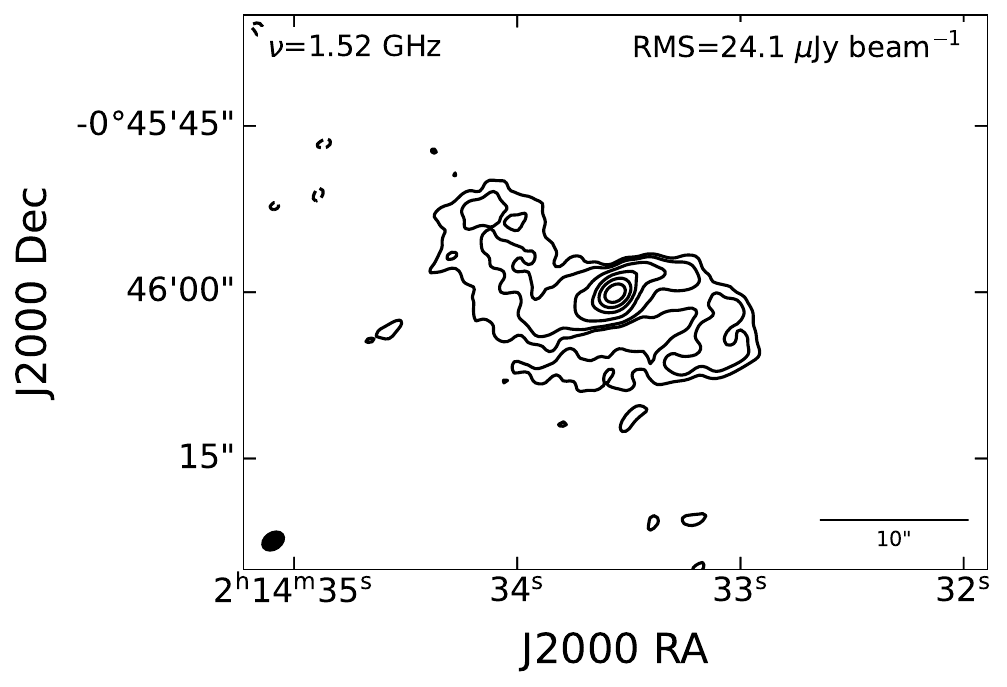}
    \includegraphics[width=\columnwidth]{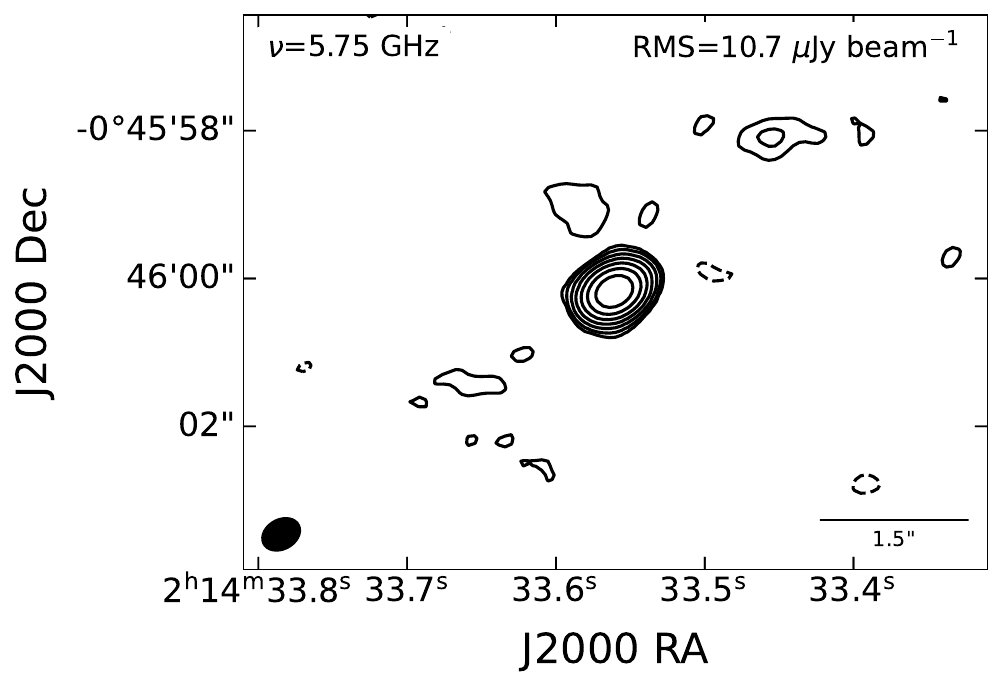}

    \includegraphics[width=\columnwidth]{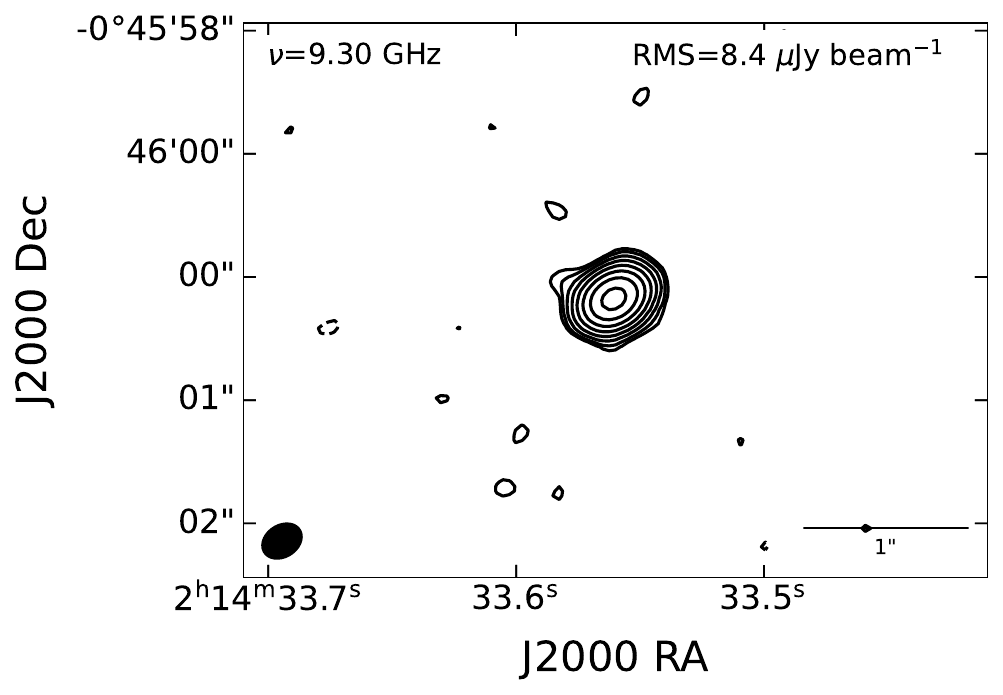}
    \includegraphics[width=\columnwidth]{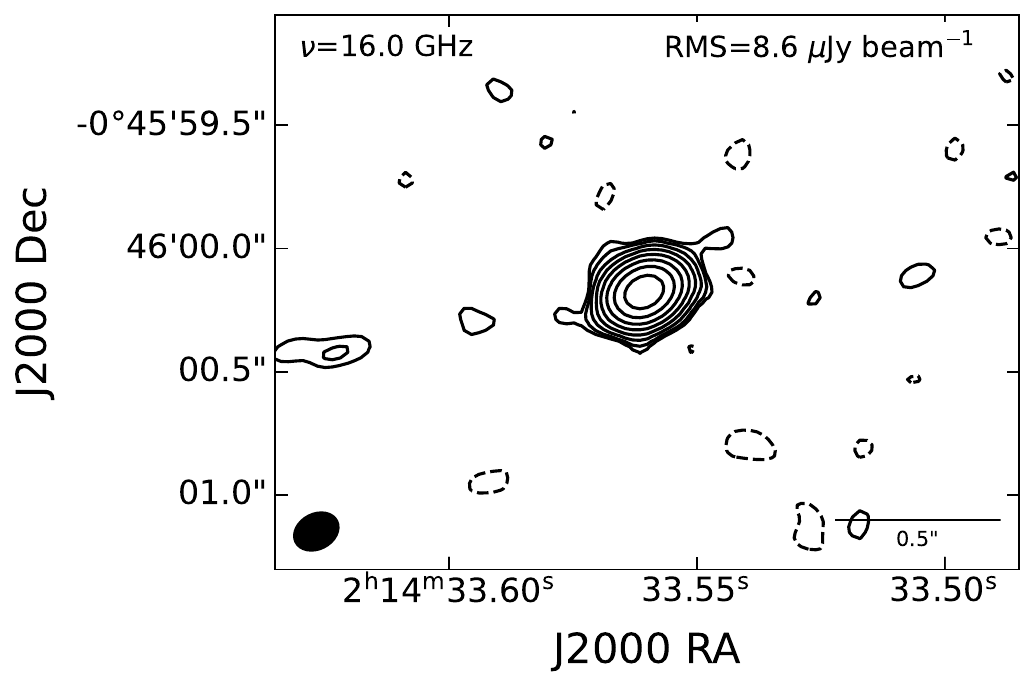}
    \caption{Intensity contour map for each stacked VLA data set. Contours start at -3,3,5$\times$image RMS and increase by a factor of 2 thereafter. The central frequency and image RMS are listed in the top left and top right for each map, respectively. The synthesized beam and a scale bar are provided in the lower left and right, respectively. For each map, north is up and east is to the right.}
    \label{fig:vla_contours}
\end{figure*}

We detect a radio source associated with Mrk 590 at each observing frequency with the VLA. The significance of detection of the peak flux density at each observing frequency exceeds $100\sigma$, where $\sigma$ is the image RMS. We present intensity contour maps of the stacked images in Figure~\ref{fig:vla_contours}. For each image, we derive the radio source parameters by performing a 2D Gaussian fit in the image plane using the CASA task \verb|imfit|. We provide the peak and integrated source flux density, image RMS and restoring beam size for the single-epoch and stacked images in Table~\ref{tab:vla_seimage_params} and Table~\ref{tab:vla_stimage_params}, respectively. The reported flux density errors are a quadrature sum of the image RMS and a 3\% calibration uncertainty in the absolute flux density scale \citep[][]{Perley&Butler_17}.

At 1.52~GHz, the radio source notably features resolved, diffuse emission with an angular diameter of $\approx20\arcsec$ (11 kpc) surrounding the central, brightest feature (Figure~\ref{fig:vla_contours}). The steepness of the diffuse component's in-band spectral index $\alpha_\mathrm{IB}\approx-0.9$ indicates an origin from optically-thin, non-thermal synchrotron emission \citep[e.\,g., ][]{Condon_92}. This feature is also spatially coincident with molecular gas rings identified by \citet{Koay+16a} and is thus most confidently associated with supernova remnants within the star-forming rings. We fit only the central, unresolved component at 1.52~GHz with the aim to exclude the flux density contribution from the diffuse, star-formation component. We report these values in Table~\ref{tab:vla_seimage_params} and Table~\ref{tab:vla_stimage_params}. We exclude flux densities for $\nu<2$~GHz from later analyses in this paper because it is difficult to confidently disentangle the star formation contribution to the radio emission from other emission process that may be associated with, e.\,g., the AGN.

At rest-frame frequencies $>2$~GHz, the source morphology is dominated by a compact component. From the 16.0~GHz image, the upper limit to the size of the unresolved component is 64~pc. At 9.30 and 16.0~GHz, the extended features to the compact component are likely imaging artifacts owing to the poor short baseline coverage with the extended VLA A-configuration. This is evident by the symmetry of the features in the 16.0~GHz image (Figure~\ref{fig:vla_contours}). However, at 5.75~GHz, we detect a diffuse component in the northeast direction at a significance of $3\sigma$ with an integrated flux density of $203~\mu$Jy. The feature is separated by 1.2\arcsec from the dominant compact compact, or 630~pc at Mrk 590's redshift, at a position angle of 17\degr. This feature is spatially coincident with a known extended \oiii feature \citep[][]{Raimundo+19} and at a similar position angle to a radio outflow detected at VLBI scales \citep[][]{Yang+21}. We discuss this feature further in Section~\ref{sec:shock} and Section~\ref{sec:corr_var}.

The in-band core spectral index $\alpha_\mathrm{IB}^\mathrm{core}$ significantly changes between all frequency pairs (Table~\ref{tab:vla_stimage_params}). At 1.52 GHz, the core in-band spectral index $\alpha_\mathrm{IB}^\mathrm{core}=-0.39\pm0.07$ approaches the demarcation between flat- and steep-spectrum sources ($\alpha=-0.5)$. 
The spectrum turns over above 1.52~GHz, where $\alpha_\mathrm{IB}>0$ and continues to rise even at 9.30~GHz. This indicates that the dominant radio emission progenitors for the low- ($\nu<2$~GHz) and high-frequency ($\nu>2$~GHz) spectra are decoupled, as expected if the low-frequency spectrum is due to star-formation processes. The core spectral index is flat by 16.0~GHz ($\alpha_\mathrm{IB}^\mathrm{core}=0.06\pm0.02$). 
The behavior of the high-frequency spectrum is characteristic of the gigahertz peaked spectrum \citep[GPS;][]{ODea&Saikia_21} sources. These are thought to be manifestations of recently-activated radio AGN. GPS sources can show spectral evolution over short time scales ($\lesssim\mathrm{years}$) depending on the physics of the injection event. 

Next, we utilize the time-series radio spectra of Mrk 590 to identify if the source has evolved over the duration of our observing campaign or if its GPS nature changed over this time. 

\begin{table*}[t!]
\centering
\caption{Mrk 590 VLA Stacked Image and Source Parameters}
\begin{tabular}{ccccccc}
\hline \hline
\noalign{\smallskip}
Band & $\nu_\mathrm{cent}$ & $S_\nu^\mathrm{peak}$ & $S_\nu^\mathrm{int}$ & $\alpha_\mathrm{IB}^\mathrm{core}$ & Image RMS & $\theta_\mathrm{maj} \times \theta_\mathrm{min}$ \\
 & (GHz) & (mJy beam$^{-1}$) & (mJy) & & ($\mu$Jy beam$^{-1}$) & (\arcsec) \\
(1) & (2) & (3) & (4) & (5) & (6) & (7) \\
\noalign{\smallskip}
\hline \\
\smallskip
L & 1.52 & 2.98$\pm$0.30 & 4.66$\pm$0.30 & -0.39$\pm$0.07 & 24.1 & 1.85$\times$1.31 \\
\smallskip
C & 5.75 & 3.31$\pm$0.33 & 3.35$\pm$0.33 & 0.16$\pm$0.03 & 10.7 & 0.47$\times$0.35 \\
\smallskip
X & 9.30 & 3.49$\pm$0.35 & 3.57$\pm$0.35 & 0.22$\pm$0.01 & 8.5 & 0.30$\times$0.22 \\
\smallskip
Ku & 16.0 & 4.63$\pm$0.46 & 4.73$\pm$0.46 & 0.06$\pm$0.02 & 8.6 & 0.16$\times$0.12 \\
\noalign{\smallskip}
\hline
\end{tabular}
\tablefoot{Frequency band (column 1) and central frequency (2) of the stacked images from our VLA observing programs of Mrk 590. For each stacked image, we also report the peak (3) and integrated flux densities (4) and in-band spectral indexes (5) of Mrk 590, and the image RMS (6) and the restoring beam (6). The reported values for 1.52~GHz correspond to the central, unresolved component of the radio source. The stacked Ku-band image excludes the single-epoch observation from program 15A-084 due to the source's flux variability, which may have affected the source morphology.}
\label{tab:vla_stimage_params}
\end{table*}

\section{Variability and Modeling of Time-Series Radio Spectra} \label{sec:modeling}
We turn to the time-series radio spectra to examine if the complex spectral behavior of the stacked radio source is consistent over the course of our VLA programs or if the source spectrum changed. As shown in Figure~\ref{fig:VLA_spectra}, the strength and shape of the broadband radio spectrum of the VLA source varies significantly epoch-to-epoch, particularly above 9 GHz. Between 2015 June 23 and 2016 October 5, the 15 and 17~GHz flux densities both increased by a factor of 2. By 2017 January 11, all flux densities above 2 GHz had significantly varied by a factor of at least 1.3. This indicates rapid variability on time scales less than 1.5~years. 

The shape of the high-frequency ($\nu>2$~GHz) spectrum is initially convex before evolving to be purely rising during the brightening phase. The peaked shape of the convex spectrum is characteristic of an absorption process. This is commonly observed for other compact radio AGN \citep[][]{ODea&Saikia_21} and can provide constraints on either the intrinsic source, such as the magnetic field strength powering the synchrotron emission, or the properties of the absorbing medium. To investigate the absorption mechanism in Mrk 590, we model the time-series, high-frequency ($\nu>2$~GHz) radio spectra with four distinct absorption processes. Our aims are to (i) constrain the best-fitting absorption model and (ii) identify if the best-fit absorption model changes from epoch-to-epoch.

\subsection{Absorption Models} \label{sec:abs_models}
The absorption models we employ for our spectral fitting routine can be broadly separated into two underlying physical mechanisms: synchrotron self-absorption (SSA) and free-free absorption (FFA). We utilize a single SSA model in our analysis. For FFA, we utilize three separate models: internal FFA, external FFA, and FFA resulting from variable ionized particle density along the line of sight. Here, we describe each absorption model we use.

\subsection*{Synchrotron Self-Absorption} \label{sec:ssa}
The spectral turnover produced by SSA occurs because the brightness temperature of the radio source cannot exceed the plasma temperature of the nonthermal electrons \citep[][]{Kellermann_66}. The spectral peak is the frequency at which the relativistic electrons and the emitted synchrotron photons have large scattering cross sections. This cross section increases with wavelength, which provides the characteristic low-frequency optically-thick spectral index of $\alpha=5/2$ below the turnover frequency for a homogeneous medium. The optically-thick emission originates from a thin shell of material surrounding the synchrotron source, while the optically-thin emission probes the interior of the source itself. If the underlying electron energy distribution is characterized by a power-law with an index of $\beta$, the observed spectral index becomes $\alpha= -(\beta-1)/2$ and the spectrum can be modeled as
\begin{equation}\label{eqn:ssa}
    S_\nu = S_0\left(\frac{\nu}{\nu_\mathrm{p}} \right)^{-(\beta-1)/2}\left(\frac{1-e^{-\tau}}{\tau} \right)\, ,
\end{equation}
where b, $\tau=(\nu/\nu_\mathrm{p})^{-(\beta+4)/2}$ is the optical depth, $S_0$ is the peak flux density, and $\nu_\mathrm{p}$ is the frequency at which the spectrum becomes optically thick \citep[][]{Kellermann_66}.

\subsection*{Free-Free Absorption} \label{sec:ffa}
FFA occurs when the synchrotron source is covered by or mixed in with an ionized screen along the line-of-sight. 
For a homogeneous medium external to the synchrotron source (EFFA), the spectrum takes the form
\begin{equation}\label{eqn:effa}
    S_\nu = S_0\, \nu^\alpha e^{\tau_\nu}\, ,
\end{equation}
where $S_0$ and $\alpha$ are the peak flux density and spectral index of the intrinsic synchrotron emission, respectively. The optical depth is $\tau_\nu = (\nu/\nu_\mathrm{p})^{-2.1}$, where $\nu_\mathrm{p}$ is the frequency at which $\tau_\nu =1$ \citep[][]{Kellermann_66}.

The thermal electrons of the ionized screen may also be mixed with the relativistic electrons of the synchrotron source. For this internal FFA (IFFA), the spectrum can be modeled as
\begin{equation}\label{eqn:iffa}
    S_\nu = S_0\, \nu^\alpha \left(\frac{1-e^{-\tau_\nu}}{\tau_\nu} \right)\, ,
\end{equation}
where the prescription for $\tau$ is identical to the EFFA case \citep[][]{Kellermann_66}.

The final absorption model that we consider accounts for an inhomogeneous density of thermal electrons in the ionized screen (InFFA). \citet[][hereafter \citetalias{Bicknell+97}]{Bicknell+97} consider the effects of an expanding jet-powered radio lobe that shocks and photoionizes a dense medium with a hydrogen number density that decreases as a power law with radius. The variable density of the medium in turn affects the optical depth of the ionized screen. \citetalias[][]{Bicknell+97} characterize the optical depth as a power-law distribution $a^p$ such that $a^p \propto \int (n_\mathrm{e}^2 T_\mathrm{e}^{-1.35})^p dl$. Implicit in their model is that the power-law index $p$ for the optical depth distribution must be $>-1$, and the scale length of the shock and/or inhomogeneities in the medium must be smaller than that of the expanding lobe. Then, the spectrum can be modeled as
\begin{equation}\label{eqn:effainh}
    S_\nu = S_0(p+1)\left(\frac{\nu}{\nu_\mathrm{p}}\right)^{2.1(p+1)+\alpha} \gamma\left[p+1, \left(\frac{\nu}{\nu_\mathrm{p}}\right)^{-2.1}\right]\, ,
\end{equation}
where $\gamma$ is the incomplete gamma function of order $p+1$.

\begin{figure}[t!]
    \centering
    \includegraphics[width=\linewidth]{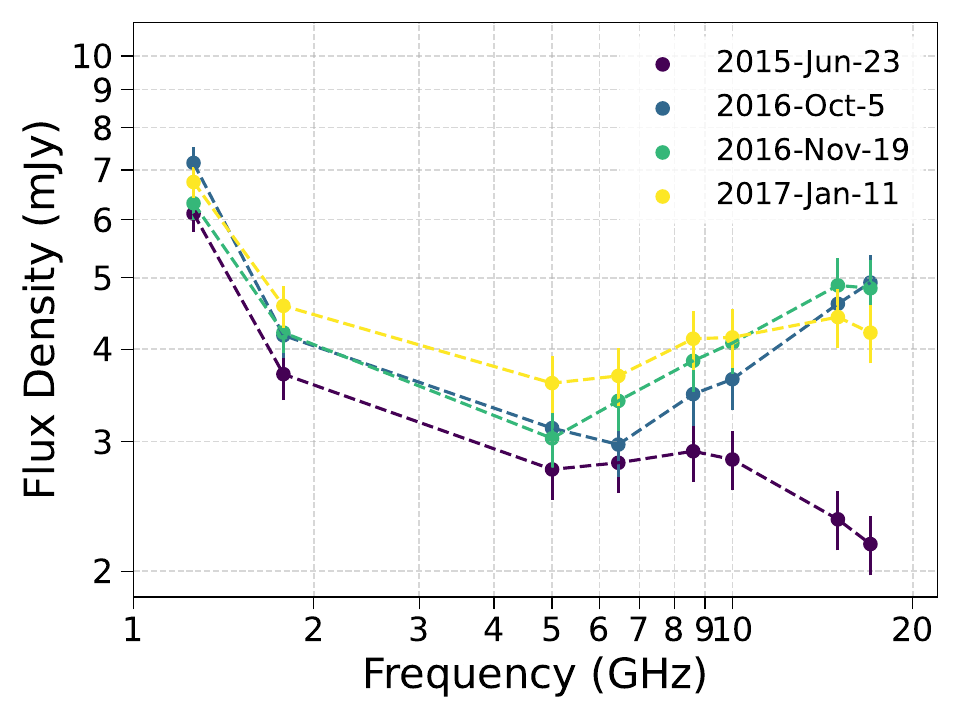}
    \caption{Time-series radio spectra of Mrk 590. All of the observations used for the spectra were conducted with the VLA in A-configuration, ensuring us that source confusion nor observational effects are not affecting the recovered flux densities. Shown for each flux density is the $3\sigma$ uncertainty, which is the quadrature sum of a 3\% uncertainty in the flux density scaling and the image RMS. (Table~\ref{tab:vla_seimage_params}).}
    \label{fig:VLA_spectra}
\end{figure}

\subsection{Bayesian Inference Modeling} \label{sec:bayesian_modeling}
We model the single-epoch, broadband radio spectra of Mrk 590 to constrain any absorption mechanism and estimate its physical parameters. We implement a Bayesian inference routine to determine the best fit model parameters and their uncertainties for each absorption model and to identify the statistically favored model for each epoch. Each absorption model requires fitting for at least three model parameters. To better constrain our model fits, we first image each spectral window of the observing bands separately, providing us with 16 flux density measurements per band. 

For our model fitting, we use a maximum likelihood estimation (MLE) to determine the most probabilistic set of model parameters given our data sets. After generating the likelihood function for each model, we perform the MLE by minimizing the negative of the likelihood function. For each model, we use a set of physically-informed bounds for our MLE: for example, the peak flux density $S_0$ and the peak frequency $\nu_p$ for each model must be positive. We then implement an affine invariant Markov Chain Monte Carlo (MCMC) ensemble sampler \citep[\texttt{emcee};][]{emcee_13} to determine the uncertainties on our model parameter fits. Each walker in the sampler is initialized by sampling a small Gaussian distribution around the maximum likelihood estimate for that parameter. We use the same physically-motivated boundary conditions for the MLE as the priors for the MCMC to constrain the sampling space for the walkers. Because the posterior distributions are sensitive to the priors, we use the same priors for each epoch when performing our inference modeling routine. We ensure that enough samples were run in the chains ($n>2\times10^5$) to reach convergence for each parameter in the model by setting a minimum number of samples threshold of at least 50 times the integrated autocorrelation time.

MLE and Bayesian MCMC routines provide the best-fit model parameters and their associated uncertainties. However, they do not provide a robust method for model selection. For model selection, we determined the Bayesian evidence for each model, given as
\begin{equation}\label{eqn:bayesian_ev}
    Z = \int \mathcal{L}(\theta)~ \Pi(\theta)~ d\theta\, ,
\end{equation}
where $\mathcal{L}(\theta)$ is the likelihood for a model parameter set $\theta$ and $\Pi(\theta)$ are the priors. The integral is over the entire dimensional space of the model parameters, and penalizes more complex models that contain a higher number of parameters. To compute the Bayesian evidence for each model, we employ the nested sampling routine \texttt{dynesty} \citep[][]{Speagle_20,Skilling_04,Higson+18}. Nested sampling evaluates the likelihoods for a global set of model parameters determined by the priors, then contracts the sampling space by discarding model parameter sets with the lowest likelihoods. Unlike \texttt{emcee}, \texttt{dynesty} provides both the posterior probability density for each model parameter and the evidence. For consistency, we check that the 50th percentile model parameter estimates determined by \texttt{dynesty} are within the error margins of those determined by \texttt{emcee}. This assures us that the computed evidences are indeed representative of the best fit model.

The evidence can then be used to determine model selection. We do this following the Jefferys scale \citep[e\,,g.,][]{Kass01061995}. For two competing models $M_\mathrm{h}$ and $M_\mathrm{l}$, with evidences $Z_\mathrm{h}$ and $Z_\mathrm{l}$, respectively, and $Z_\mathrm{h} > Z_\mathrm{l}$, the model selection criterion in log space ($\Delta \mathrm{ln}(Z) = \mathrm{ln}(Z_\mathrm{h}) - \mathrm{ln}(Z_\mathrm{l})$) is: $0\leq\Delta \mathrm{ln}(Z)\leq1.16$ is inconclusive; $1.16<\Delta \mathrm{ln}(Z)\leq2.3$ is moderate evidence in favor of $M_\mathrm{h}$; and $\Delta \mathrm{ln}(Z)>2.3$ is strong evidence in favor of $M_\mathrm{h}$. As a consistency check, we also determine the Bayesian information criterion \citep[BIC;][]{BIC} for each model fit. We evaluate the BIC for each model fit by using the MLE parameter estimates. We note that we do not use the BIC as the primary model selection criterion because it is a large-sample size approximation to the more general Bayes factor and the population of data points in our fit is only a factor of at most 16 larger than the number of model parameters.

\begin{figure*}
    \centering
    \includegraphics[width=\columnwidth]{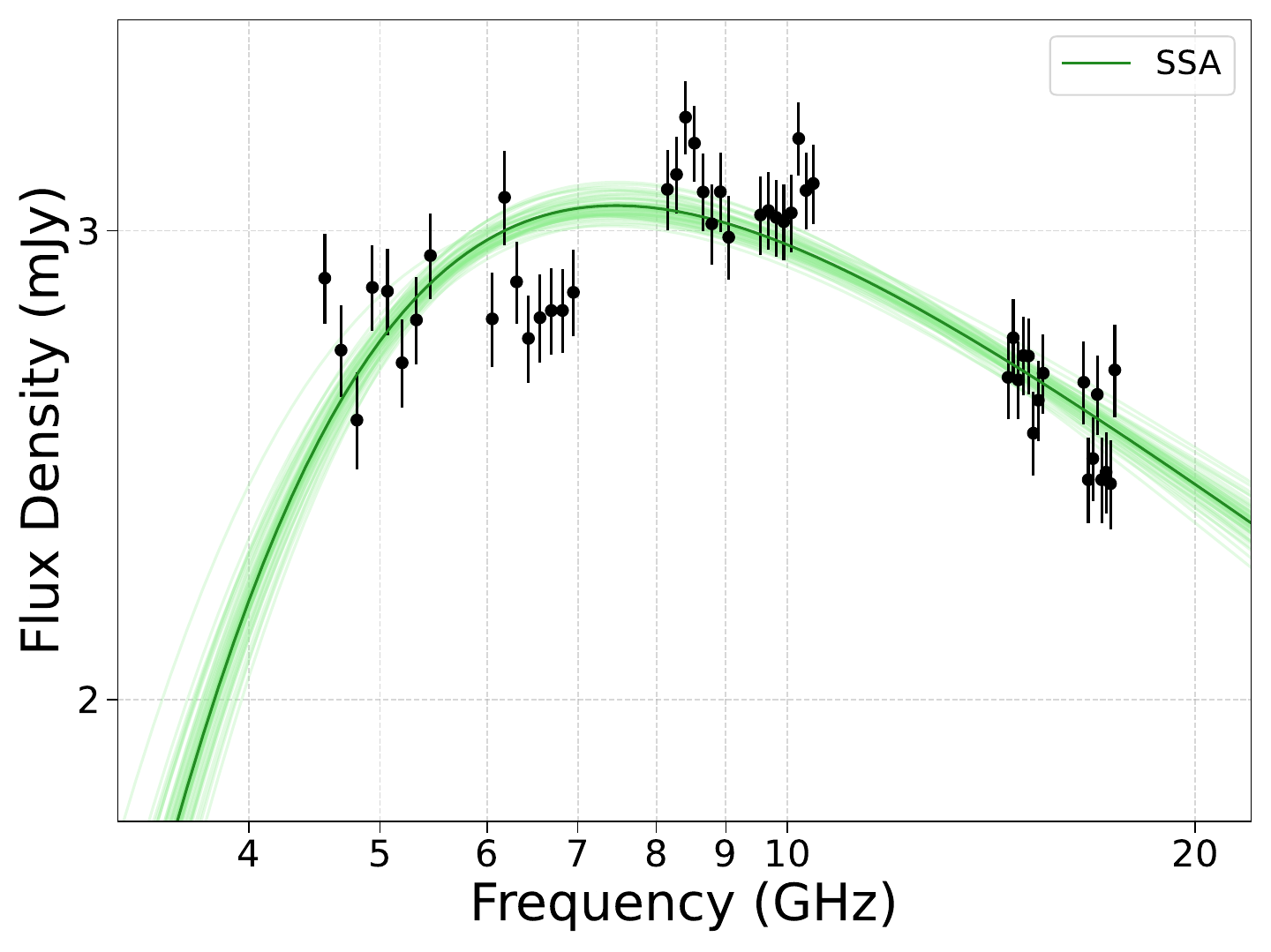}
    \includegraphics[width=\columnwidth]{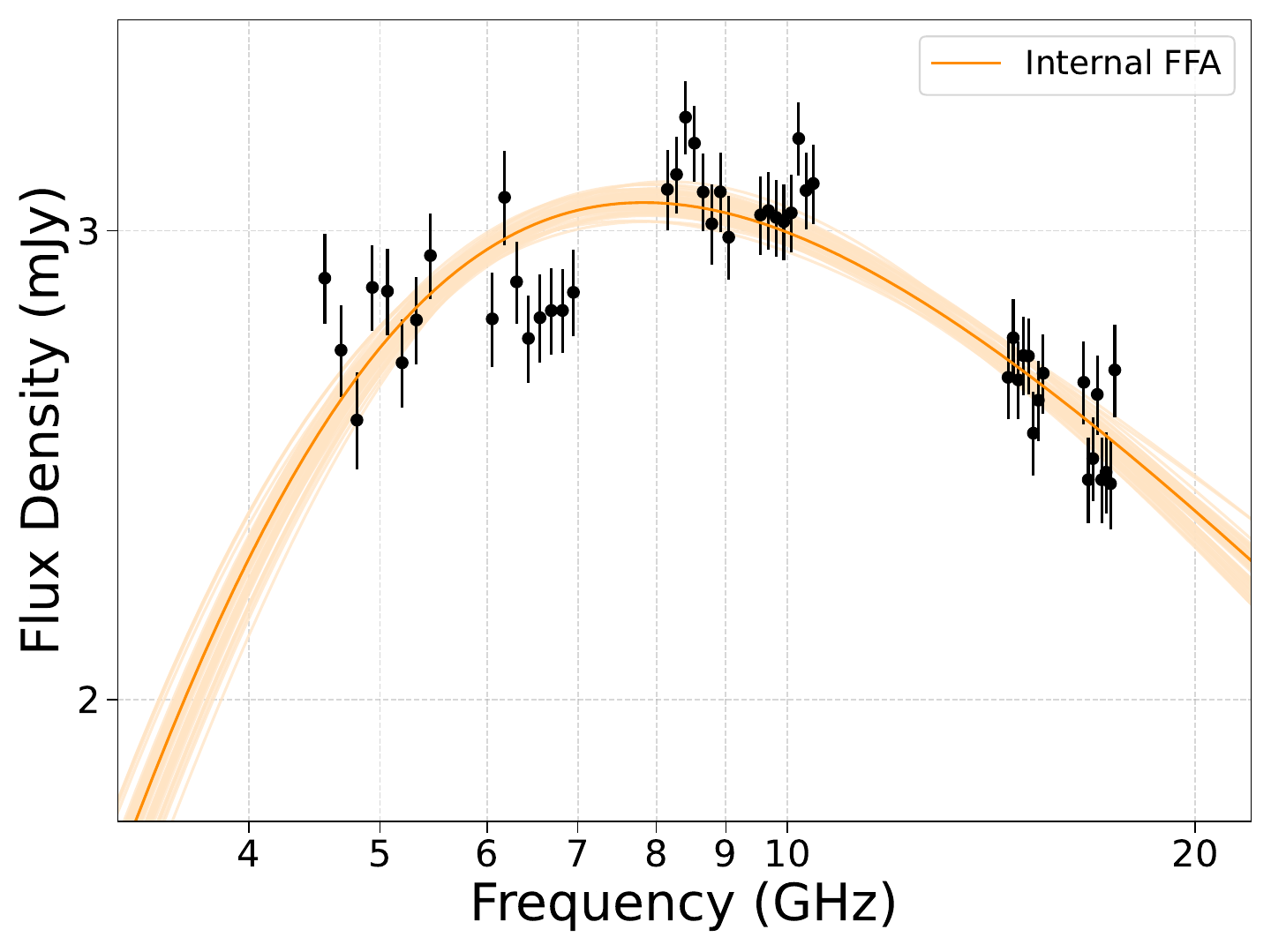}

    \includegraphics[width=\columnwidth]{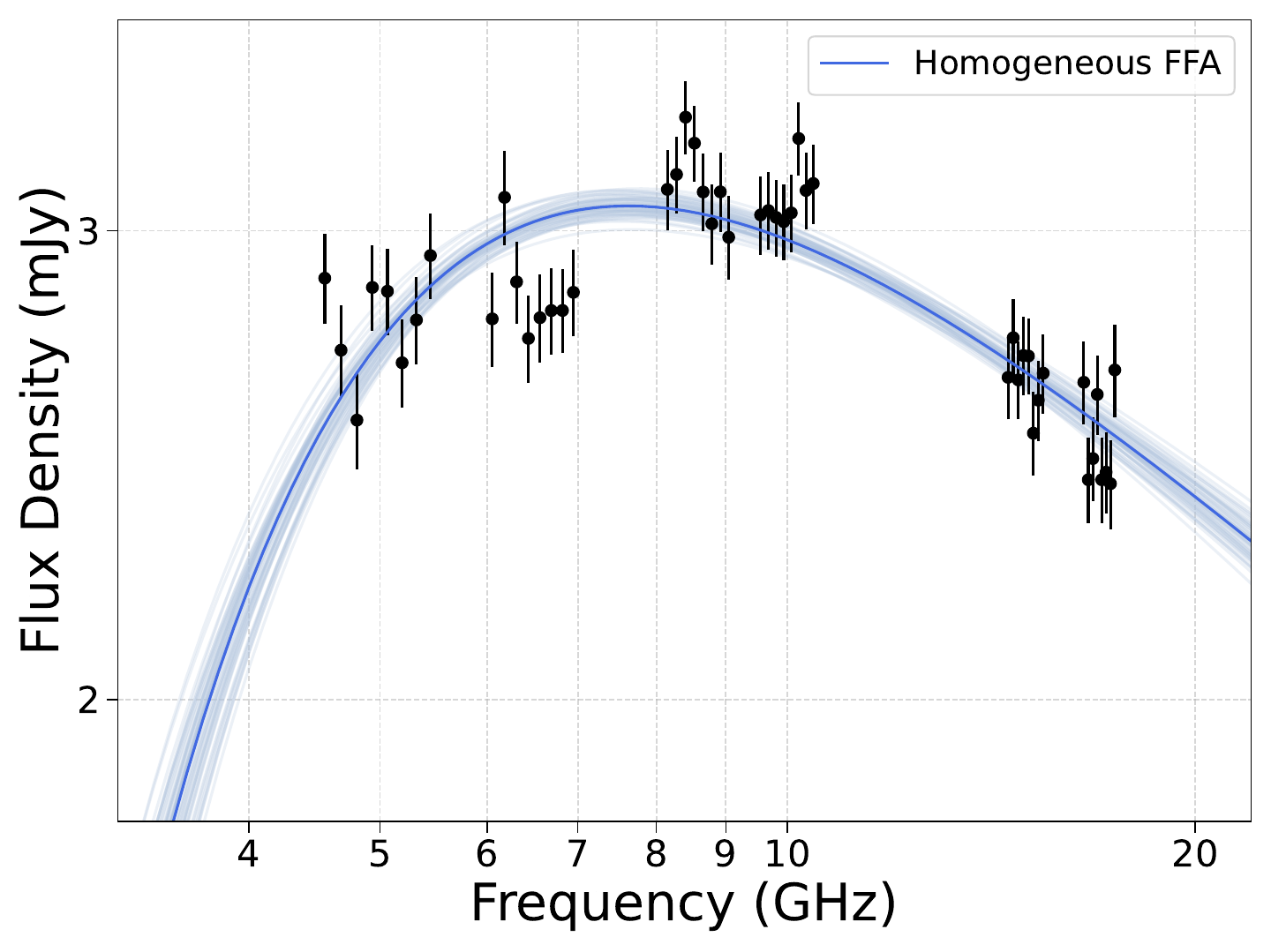}
    \includegraphics[width=\columnwidth]{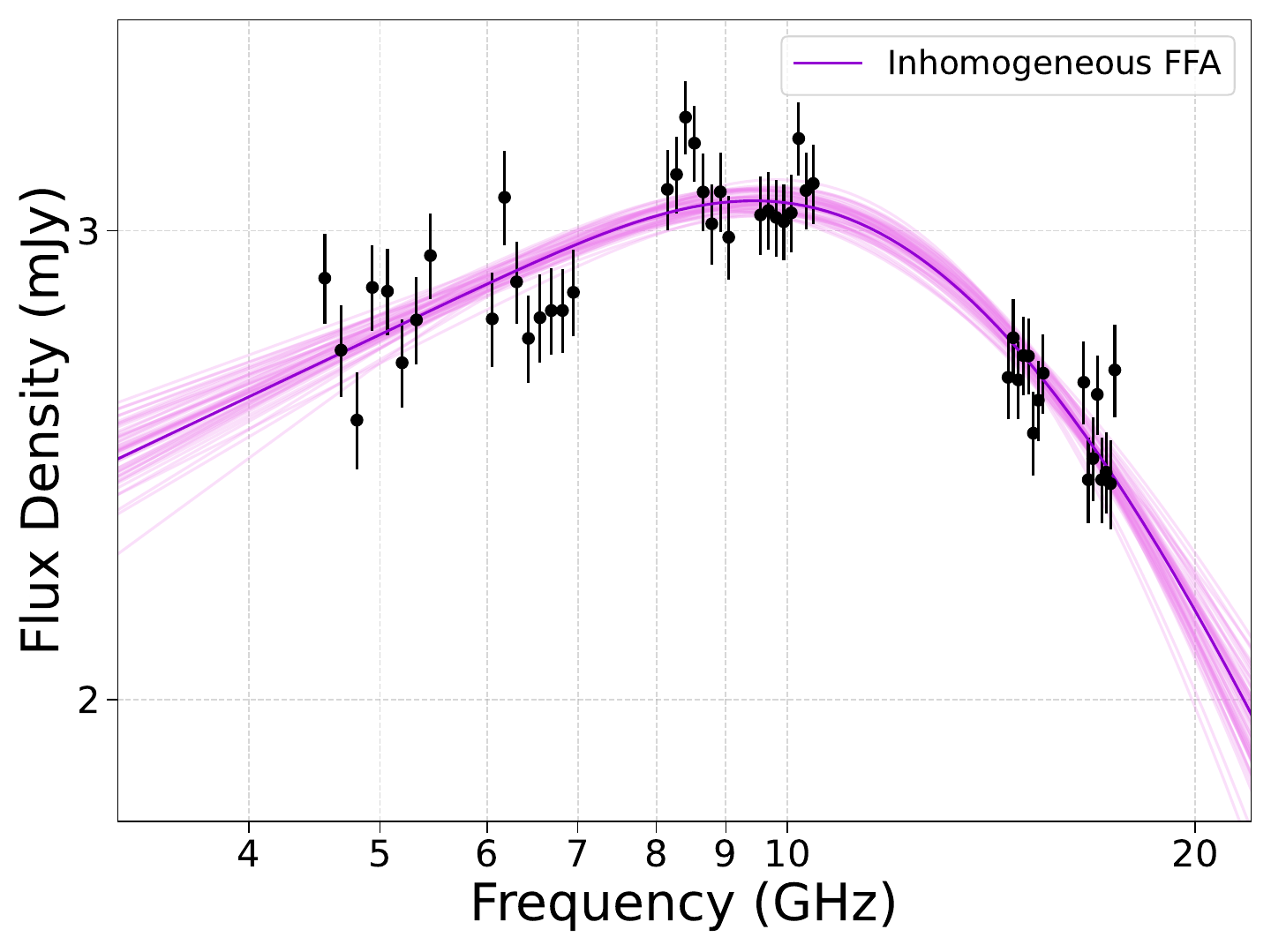}
    \caption{Absorption model fits to the epoch 1 (2015 June 23) VLA radio spectrum of Mrk 590. We list the best fit model parameters and their uncertainties in Table~\ref{tab:model_results}. Flux density measurements from our VLA observations are shown as black data points, with errors that are a quadrature sum of a 3\% uncertainty in the flux density scale and the image RMS. The maximum likelihood parameter fit to the data is shown as the darkest line in each spectrum, with 50 randomly-drawn individual realizations of our Bayesian MCMC shading this fit.}
    \label{fig:model_fits}
\end{figure*}

Our inference modeling is only able to distinguish between different absorption models for the epoch 1 (2015 June 23) data set, following the Jefferys scale for $\Delta\mathrm{ln}(Z)$. We summarize the results of our model fits and inference selection statistics for epoch 1 in Table~\ref{tab:model_results}, and we present the absorption model fits to the epoch 1 radio spectrum in Figure~\ref{fig:model_fits}.
The inability to determine the best-fit model for epochs 2 (2016 October 05), 3 (2016 November 19) and 4 (2017 January 10), which correspond to the radio brightening period of Mrk 590, is due to the lack of frequency coverage above 18~GHz, where we expect the spectral peak frequency to be for these latter epochs. To fit any of the given absorption models requires resolving the spectral peak in frequency space. Even by eye (Figure~\ref{fig:VLA_spectra}), our observations do not resolve the peak frequency for epochs 2 and 3. To test this for epoch 4, we fit this radio spectrum with a phenomenological model that describes a parabola in log space. We do not identify significant curvature of this spectrum. This is in agreement with the indetermination of a favored model from our inference modeling.

From our inference modeling of epoch 1 (Table~\ref{tab:model_results}), the Bayesian evidence (minimum $\Delta \mathrm{ln}(Z) = 12.5$) and BIC (minimum $\Delta\mathrm{BIC=23.6}$) both suggest that the inhomogeneous FFA model (Eqn.~\ref{eqn:effainh}) is the preferred description for the convex shape of the radio spectrum. Our inference modeling informs us that over the duration of our observing campaign (2015 June - 2017 January), Mrk 590's radio spectrum always had a peak frequency above 10~GHz, with the inhomogeneous FFA model providing the highest $\nu_\mathrm{p}$ of any model fit to the epoch 1 data set. During our epochs of observation, Mrk 590 was always a Gigahertz-peaked spectrum (GPS) source. This is notable in that GPS sources are historically thought to be young radio AGN that have launched jets within the last several thousand years \citep[][]{ODea&Saikia_21}. If this applies to Mrk 590, its GPS nature both in 2015 and 2016-2017 could indicate the recent launching of new jetted components that may be correlated with the accretion state transition of 2017. We discuss this in Section~\ref{sec:corr_var}. 

\begin{table*}[t!]
\centering
\caption{Inference Modeling Parameters and Model Selection Criterion for Epoch 1}
\begin{tabular}{lccccccc}
\hline \hline
\noalign{\smallskip}
Model & $S_0$ & $\nu_\mathrm{p}$ & $\alpha$ & $\beta$ & $p$ & $\mathrm{ln}(Z)$ & BIC \\
 & (mJy) & (GHz) & & & & & \\
(1) & (2) & (3) & (4) & (5) & (6) & (7) & (8) \\
\noalign{\smallskip}
\hline \\
\medskip
SSA & 4.14$\pm$0.07 & $4.73^{+0.15}_{-0.16}$ & -0.37$\pm$0.03 & 1.74$\pm$0.03 & - & 4.0$\pm$0.30 & -26.3 \\
\medskip
Internal FFA & $11.9^{+1.66}_{-1.37}$ & $5.92^{+0.33}_{-0.32}$ & $-0.53^{+0.04}_{-0.05}$ & - & - & 12.1$\pm0.3$ & -39.3 \\
\medskip
Homogeneous FFA & $9.67^{+0.93}_{-0.84}$ & $3.63^{+0.15}_{-0.16}$ & -0.46$\pm$0.03 & - & - & 7.8 $\pm$0.3 & -32.8 \\
\medskip
Inhomogeneous FFA & $3.87^{+0.14}_{-0.19}$ & $15.60^{+2.03}_{-1.86}$ & $-1.39^{+0.27}_{-0.35}$ & - & $-0.22^{+0.15}_{-0.11}$ & 24.6$\pm$0.3 & -62.9 \\
\noalign{\smallskip}
\hline
\end{tabular}
\tablefoot{The absorption model (column 1), and the peak flux density (2) and frequency (3), and spectral index (4) model fits as determined by our MLE routine. (5) and (6) are the power law index for the electron energy distribution of the SSA model and the power law index for the distribution of optical depths for the InFFA model, respectively. Log-evidence (7) and BIC (8) for each model as determined through nested sampling.
Error margins for all model parameter estimates correspond to the 16$^\mathrm{th}$ and 84$^\mathrm{th}$ percentiles of the posterior distribution.}
\label{tab:model_results}
\end{table*}

\section{Radio Variability Mechanism} \label{sec:var_origins}
We have shown that the convex radio spectrum of Mrk 590 during our epoch 1 observations (2015 June 23) was most likely due to absorption of the underlying synchrotron emission generated by an interaction between a radio lobe and the inhomogeneous interstellar medium (ISM). We have also shown that, within 16 months, the radio spectrum evolved away from a convex shape and the radio source entered a brightening period. We now consider different models for the radio variability mechanism to assess if the variability at radio wavelengths could be correlated with the variability observed in the optical/UV/X-ray wavebands.

\subsection{Scintillation} \label{sec:scintillation}
Scintillation events are caused by the scattering or lensing of synchrotron photons as they pass through the interstellar medium of the Milky Way along our line of sight, leading to variability of the observed flux density of the source with time. However, there are several reasons why the radio brightening of Mrk 590 cannot be due to scintillation. Scintillation time scales are much shorter than the duration of the radio variability we have discovered. For mid-Galactic latitudes, such as for Mrk 590, the transition frequency between the strong and weak scattering regimes is $\nu_o\approx8$~GHz \citep[][]{Walker_98, Walker_01}. For this exercise, we only consider the observed frequencies of Mrk 590 that fall within the weak scattering regime.
In the weak regime, the time scale of the flux modulation is $t\sim2(\nu_o/\nu)^{1/2}$ in hours. For $\nu\geq8$~GHz, flux modulation due to scintillation cannot exceed $t=2$~hours. We observe a persistent enhancement of Mrk 590's broadband flux densities for months, thus ruling out scintillation as the variability mechanism. Additionally, the broadband variability amplitudes we identify ($\approx100$\%) are too large to be explained by weak scintillation, which can produce variability amplitudes of at most 10\%. The variability mechanism must be source intrinsic and rapid, having evolved on time scales of only $\approx1.5$~years.

\subsection{Tidal Disruption Event} \label{sec:tde}
Tidal disruption events (TDEs) are produced when a star is ripped apart by the gravitational potential of a supermassive black hole. The multi-waveband behavior of TDEs traces different physical processes of the event: optical/UV and X-ray emission is thought to trace the mass fallback and cooling of the disrupted material \citep[][]{Stone+13,Metzger_22}; radio emission traces the outflows, including relativistic jets, launched after the formation of the accretion disk \citep[][]{Cendes+24}. In radio-detected TDEs, the radio emission is ubiquitously observed to follow the optical/UV/X-ray brightening \citep[][]{Cendes+24,Alexander+26}. In Mrk 590, we observe the initial radio brightening to precede the optical/UV/X-ray brightening by at least a year. If Mrk 590 were a dust-obscured TDE at early times such that the optical/UV/X-ray emission were attenuated to levels below detection thresholds, this would necessitate a time-varying line-of-sight column density. However, modeling of Mrk 590's 2013 X-ray spectrum, while the AGN was in its historically low flux state, found no evidence for an obscuring medium \citep[][]{Denney+14}. We can definitively rule out a TDE as the radio brightening mechanism for Mrk 590.

\subsection{Shock Front} \label{sec:shock}
As a radio jet propagates outwards after collimation in the vicinity of the SMBH, the jet drives the expansion of radio lobes. These lobes interact with dense reservoirs of gas in the ISM of the host galaxy, driving a shock front at the interface of this interaction. The shock front accelerates electrons along compressed magnetic field lines, producing a new population of synchrotron photons that cause a brightening of the associated radio source. If the energetics of the shock front are sufficiently powerful, these electrons can generate synchrotron radiation at rest-frame frequencies in the GHz regime, producing GPS sources. Additionally, the shock front ionizes the ISM, producing optical line emission. In this subsection, we investigate a shock front as the mechanism driving the radio variability we identified in Mrk 590. 

First, we investigate the spatial scale at which this variability mechanism may be occurring to understand if our observations had the ability to resolve the length scale of the interaction. From our multi-epoch imaging, we do not identify any new components in the single-epoch images of Mrk 590. This suggests that the scale size of the variability mechanism is smaller than we can probe with our VLA observations. We first constrain the scale size of the lobe-ISM interaction from epoch 1 using our $\nu_\mathrm{p}$ estimate for the inhomogeneous FFA model (Table~\ref{tab:model_results}) and the empirical turnover$-$size relation of \citet[][]{ODea&Baum_97}. The turnover$-$size relation is a consequence of SSA or FFA in the lobes of a young radio jet. As the source expands, the opacity decreases, shifting the spectral peak to progressively lower frequencies. This is also reproduced by the inhomogeneous FFA model of \citetalias[][]{Bicknell+97}. The linear size follows $LS\propto\nu_\mathrm{p}^{1.65}$ \citep[][]{ODea&Baum_97}.

The intrinsic source size estimated from this relation is 7~pc. The minimum resolvable scale in our highest resolution image is 50~pc. If the synchrotron source is due to a bow shock produced through a jet-ISM interaction, it cannot have occurred more than $\sim45$~pc from the known VLBI radio core \citep[][]{Koay+16b}. If the bow shock was produced by an interaction further than this distance, the 16.0~GHz source morphology (Figure~\ref{fig:vla_contours}) would likely be resolved, at least marginally. In our later epochs while the source was brighter, we cannot directly estimate the source size because we cannot constrain the peak frequency. However, if we assume $\nu_\mathrm{p} =$ 17.4~GHz for the three later epochs, which is the highest central frequency of any individual spectral window in our observations, the upper limit to the source size is 6~pc. 
Given the small inferred linear sizes of the variable radio component with respect to the synthesized beam of the VLA, we conclude that it is unlikely that we could directly resolve this component in any of our single-epoch images.

The strongest evidence that supports a shock front as the radio variability mechanism is the persistent, broadband enhancement of Mrk 590's flux densities in the later three epochs of our program. In the propagating shock front model of \citetalias[][]{Bicknell+97} the magnetic field lines are compressed along the shock front interface. Electrons accelerate along these magnetic field lines, producing a new population of high-energy synchrotron photons that drive enhanced flux densities. In particular, \citet[][]{Tingay&Kool_03} note that the \citetalias[][]{Bicknell+97} model is expected to drive variability above the intrisnic spectral turnover frequency, which we find to be $\nu_p=15.6$~GHz through our model fit to the epoch 1 radio spectrum. In the later epochs, the strongest observed variability occurred at 15~GHz and 17~GHz. The rising in-band spectral index values ($\alpha_\mathrm{IB}>0.1$) at these frequencies during epochs 2 (2016 October 5) and 3 (2016 November 19) imply that stronger variability amplitudes occurred above these frequencies, consistent with the shock front model.

It is expected that a jet-driven shock front will produce optical ionization lines that are velocity shifted with respect to their rest frame, due to the jet propagating radially outwards into the ISM. Curiously, the broad line features of Mrk 590, which disappeared between 2006 and 2012 \citep[][]{Denney+14}, re-appeared by late 2017. \citet[][]{Raimundo+19} identified a broad (FWHM~$=1650$~km~s$^{-1}$), blue-shifted ($V\approx-400$~km~s$^{-1}$ with respect to the rest frame, $6\sigma$ significance) component to the narrow \oiii emission line in the integrated $8\arcsec \times 8\arcsec$ nuclear spectrum with MUSE observations in 2017. We utilize the \citetalias[][]{Bicknell+97} model to investigate if the shock front that drove the radio variability also produced the outflowing \oiii emission.

From \citetalias[][]{Bicknell+97}, the \oiii luminosity produced by the shock can be determined analytically as
\begin{equation}\label{eqn:ionized_shock}
    L_{\oiii} = 10^{43}\left(\frac{6}{8-\delta} \right) \left(\frac{\kappa_\nu}{10^{-11}} \right)^{-1} \left(\frac{P_\nu}{10^{27}\ \mathrm{W\ Hz^{-1}}} \right)\ \mathrm{erg\ s^{-1}}\, ,
\end{equation}
where $\delta$ is the power-law index describing the density profile of the ISM, $\kappa_\nu$ is the conversion from jet energy flux to monochromatic radio power, and $P_\nu$ is the monochromatic radio power. For the epoch 1 fitted model, following \citetalias{Bicknell+97} to adopt $\delta=2$ and vary $\kappa_\nu$ between $10^{-10.5}$ and $10^{-11.5}$, the \oiii luminosity produced by the bow shock is between $2\times10^{37}$~erg~s$^{-1}$ and $2\times10^{38}$~erg~s$^{-1}$.

We fit the integrated intensity map of the broad \oiii component from \citet[][]{Raimundo+19} with a 2D Gaussian to determine its total flux. We find $F_{\oiii}=9.6\times10^{-15}$~erg~s$^{-1}$~cm$^{-2}$, or a luminosity of $L_{\oiii}=1.6\times10^{40}$~erg~s$^{-1}$. This is nearly two orders of magnitude higher than what is expected if only the bow shock powered the blue-shifted emission. The central engine of the AGN itself must be contributing to the energetic budget of this component. However, closer inspection of the integrated intensity map shows a radially-extended component to the blue-shifted \oiii emission, extending 750~pc north-south from the nucleus \citep[Figure~10 of][]{Raimundo+19}. To determine the luminosity of this component, we first subtracted the 2D Gaussian model fit to the integrated intensity map. We then fit the residuals with a separate 2D Gaussian and find a total flux to the extended component of $F_{\oiii~\mathrm{Ext.}}=1.2\times10^{-15}$~erg~s$^{-1}$~cm$^{-2}$, or a luminosity of $L_{\oiii~\mathrm{Ext.}}=2\times10^{39}$~erg~s$^{-1}$. This is still one order of magnitude brighter than our estimated upper limit to the \oiii luminosity produced by the bow shock. 

We also test if the northeast radio component detected at 5.75~GHz (Figure~\ref{fig:vla_contours}) may be driving the \oiii outflow at 750~pc from the nucleus. Assuming this radio feature corresponds to a separate shock, we estimate the shock-driven \oiii luminosity following Eqn.~\ref{eqn:ionized_shock} and the 5~GHz monochromatic radio power of this radio feature. We find an estimated shock-driven \oiii luminosity of $L_{\oiii} = 1\times10^{37}$erg~s$^{-1}$, which is two orders of magnitude lower than the observed $L_{\oiii}$ of the 750~pc extended \oiii feature. Although the northeast radio component is spatially coincident with the extended \oiii feature, we do not find evidence to suggest that the energetics of this radio component alone are powering the \oiii emission.

An important caveat to our analysis is that, as noted by \citetalias[][]{Bicknell+97}, the value of $\kappa_\nu$, the conversion factor between jet energy flux and monochromatic radio power, is poorly constrained. However, the values we have used for $\kappa_\nu$ ($10^{-10.5}-10^{-11.5}$) were determined by considering the monochromatic radio luminosities of radio galaxies. These sources have $P_\nu$ as high as $10^{27.5}$~W~Hz$^{-1}$, whereas Mrk 590 only has $P_\nu=3.5\times10^{20}$~W~Hz$^{-1}$ from epoch 1 of our observations. This is a limiting constraint on the model we have considered, and likely leads to large variations in the estimated $L_{\oiii}$ values. For this reason, we cannot definitively rule out that the radio source powered the blue-shifted \oiii outflow.

\section{Correlated Variability with the Changing-State Transition}\label{sec:corr_var}
Our Bayesian inference modeling has shown that the most likely model producing the enhanced radio flux density variations of Mrk 590 between 2016 and 2017 is that of a bow shock that re-accelerates energetic electrons in the ISM. From the turnover-size relation (Section~\ref{sec:shock}), the upper limit to the size of the shock region is $6$~pc. To attain the scale sizes associated with the central engine of the AGN, i.\,e., the broad line region, accretion disk, X-ray corona, would require a radio source with a linear size of the order $0.01$~pc \citep[][]{Peterson+04}, and thus a peak frequency that is orders of magnitude higher than the likely peak of the epochs 2, 3, and 4 radio spectra, following \citet[][]{ODea&Baum_97}. This informs us that the scale size of the shock is much larger than the scale sizes associated with the central engine of the AGN, i.\,e., the broad line region, accretion disk and X-ray corona. The radio brightening we observed is decoupled from the optical/UV/X-ray brightening identified beginning in 2017 \citep[][]{Raimundo+19,Lawther+23}. The optical/UV/X-ray variability of CSAGN is known to correlate with a state transition of the accretion flow \citep[e.\,g.,][]{Denney+14,Mathur+18}, and thus the variability mechanism is occurring at the scale size of the accretion disk. Although the radio brightening we observed is temporally coincident with the reawakening of the AGN in 2017, the bow shock must be occurring on spatial scales much larger than the size of the accretion disk. 
The decades-long flux density behavior of Mrk 590 shows a dimming of the radio source as the AGN transitioned to a radiatively inefficient flow \citep[][]{Koay+16b}. However, our VLA observations here do not suggest a direct causal relation between Mrk 590's radio source and the 2017 accretion state transition. We will further investigate this in an upcoming paper that utilizes multi-epoch monitoring with the VLBA, which mitigates the contributions to the radio emission from the shock event.

The CLAGN 1ES 1927+654 exhibited clear evolution of its radio source years after its changing-look event \citep[][]{Meyer+25}. The extraordinary enhancement in the source's flux density and the launching of a bipolar outflow are each suggestive of dynamic magnetic fields close to the supermassive black hole \citep[][]{Laha+25}. The order of these events, with brightening of the higher energy emission preceding the radio emission, suggests a causal link between the accretion variability mechanism and the radio source. For Mrk 590, the low amplitude radio variability could imply a weak coupling between the radio source and the accretion flow. However our inference modeling is critical in providing the key evidence that any broad changes to the magnetic fields, in amplitude and/or topology, at the scale size of the central engine are insignificant throughout Mrk 590's changing-look event in 2017. This is supported by the stability of the soft X-ray excess at the low-intermediate accretion rate of Mrk 590 in 2017 \citep[$\lambda_\mathrm{Edd}$ = 0.03][]{Lawther+23}. The soft X-ray excess is thought to originate via Compton up-scattering of UV seed photons by a warm, magnetically supported coronal component \citep[][]{Czerny&Elvis_87,Done+12}. The temperature of the warm corona would vary with the magnetic fields threading the accretion disk \citep[][]{Fabian+15}. However $kT_\mathrm{warm}$ remains constant between Mrk 590's flux states \citep[][]{Lawther+25}, providing supporting evidence that magnetic field variability was insignificant during the accretion state transition. For these two CLAGNs, the driving mechanisms behind their accretion state transitions must be distinct from each other, promoting the idea that changing-state behavior is driven by a diverse class of mechanisms. 

The radio feature at 5.75~GHz with a 630~pc separation northeast from the core (Figure~\ref{fig:vla_contours}) and the presence of an outflow with a similar position angle at VLBI scales \citep[][]{Yang+21} suggests that Mrk 590 may be episodically ejecting outflowing radio components.
However, it is evident that this 5.75~GHz component is not related to the changing-look events that Mrk 590 has undergone in the past few decades. If we assume the proper motion of the 5.75~GHz, VLA component to be $v=c$, it was ejected from the VLBI core $\sim2$~kyr ago. The only possibility for this time scale to shorten is if the component were ejected closely along our line of sight. If this were true, its ballistic motion could be superluminal, as is observed for ejected components of blazars \citep[e.\,g.,][]{Lister+13}. However, the spectral energy distribution of Mrk 590 is indicative of a host galaxy contribution plus AGN accretion disk structures \citep[][]{Denney+14,Lawther+23}, unlike the characteristic double hump shape of a blazar due to the dominance of Doppler boosted synchrotron emission \citep[][]{Padovani+17}. We therefore rule out the possibility of superluminal motion for the ejected radio components of Mrk 590. More observations are needed to test the hypothesis that these radio outflows correspond to distinct transitions of the accretion flow owing to changing-look events.

\section{Conclusions}\label{sec:conclusions}
The CLAGN Mrk 590 has exhibited dramatic photometric and spectroscopic variability over the past 40 years. After entering its lowest flux state in 2014 \citep[][]{Denney+14}, Mrk 590 began to reawaken in 2017, showing clear signatures of an accretion state transition in the optical, UV, and X-ray wavebands. We carried out four epochs of high-resolution observations from 1-17~GHz with the VLA between 2015 and 2017 to assess if the radio-emitting components responded to the 2014 changing-look event. Our findings are as follows:

\begin{itemize}
    \item The dominant radio component at all frequencies is the compact core. Spatially extended features are absent except at 1.5~GHz, where synchrotron emission associated with supernova remnants is detected, and 5.75~GHz, where we detect a secondary component at a similar PA to the VLBI feature reported by \citet[][]{Yang+21}.
    \item The core shows a time-variable radio spectrum above rest-frame frequencies of $2$~GHz. The spectrum is initially convex in our 2015 epoch, indicating an absorption process, before brightening and rising between 2016 and 2017. We discovered that the maximum flux density variations were by a factor of two or less from their initial values in our 2015 epoch. Our observations do not identify a statistically significant decline in the flux densities in the latest 2017 epoch.
    \item We fit the epoch 1 (2015 June 23) high frequency ($\nu>2$~GHz) spectrum with four different absorption models: synchrotron self-absorption (SSA); external free-free absorption by a medium of homogeneous density (EFFA); internal free-free absorption (IFFA); external free-free absorption by a medium of inhomogeneous density (InFFA). We implement a Bayesian inference routine to identify the best-fit model to the epoch 1 radio spectrum. We find that the InFFA model best described our data set based on its Bayesian evidence. 
    \item The multi-frequency variability we identify in the latter three epochs cannot be due to interstellar scintillation by the ISM of the Milky Way, as the modulation time scale and variability amplitude of the radio emission are inconsistent with the expectations of weak scintillation. We also consider a TDE as the variability mechanism, however the onset of the radio brightening for Mrk 590 precedes that at the optical/UV/X-ray wavebands by at least a year, whereas radio emission from TDEs is ubiquitously observed at later times than the optical/UV/X-ray events. Variable dust obscuration cannot cause the radio-first brightening in Mrk 590 as X-ray spectral modeling has found no obscuring material present around the AGN \citep[][]{Denney+14}. 
    \item Most likely, the radio brightening observed in our 2016 and 2017 epochs is a result of a bow shock between an expanding radio lobe and the ISM of the host galaxy, as is suggested by the statistically-favored InFFA model. The bow shock compresses galactic magnetic field longs along the shock front interface, creating particle acceleration sites that produce a new population of synchrotron photons. This drove the broadband flux density enhancement we observe in the latter epochs of our program, with the highest inferred variability amplitudes occurring above the best-fit peak frequency of the 2015 June epoch. 
    \item IFU observations of Mrk 590 in late 2017 identified the presence of a blue-shifted outflow to the \oiii emission. The elongation along the north-south axis and radial extent of this \oiii outflow is consistent with the detection of an extended radio component identified at 5.75~GHz. We fit this \oiii outflow to determine its observed luminosity. We estimated the \oiii luminosities produced by the bow shock and the extended radio component separately to assess if either drive the ionized outflow. Although the estimated \oiii luminosities are one or more orders of magnitude lower than the observed luminosity, we cannot rule out that either of these components is driving the \oiii outflow, as our analysis is limited by poor constraints on the conversion factor between jet energy flux and monochromatic radio power.
    \item Mrk 590 began exhibiting dramatic optical-UV-X-ray variability in 2017, just after the radio variability we identified. However, the bow shock model we favor for the radio variability is unassociated with the accretion-driven variability identified in the higher energy wavebands, as it occurs much further downstream from the AGN itself. This also suggests that the 2017 changing-look event did not significantly alter the magnetic fields, in strength and/or topology, close to the AGN. Juxtaposing this with other CLAGN (e.\,g., 1ES 1927+654) that have demonstrated significant magnetic field variability, this promotes the case that changing-state are driven by a diverse population of mechanisms.
\end{itemize}

Although the radio variability we identify here is unassociated with the accretion-driven variability observed in the optical/UV/X-ray wavebands, we cannot explicitly argue against a causal connection between the accretion disk and radio source, as the correlated variability may be much weaker than the bow shock event. Higher resolution observations with VLBI may resolve out the emission associated with the bow shock, laying bare the radio core and permitting a closer inspection of the causal link between the radio and accretion-driven emission. In an upcoming paper, we will address the question of correlated radio variability with the changing-state mechanism, and thus the prevalence of a disk-jet connection in AGN, through analysis of a long-term monitoring program with the VLBA. 

\begin{acknowledgements}
This work is supported by the Independent Research Fund, Denmark via grant number DFF-8021-00130 and the Carlsberg Foundation via grant CF21-0649. SIR acknowledges support by the Science and Technology Facilities Council (STFC) of the UK Research and Innovation via grant reference ST/Y002644/1 and from the Royal Society via grant reference RGS\textbackslash R2\textbackslash 252734. The National Radio Astronomy Observatory and Green Bank Observatory are facilities of the U.S. National Science Foundation operated under cooperative agreement by Associated Universities, Inc. We acknowledge the use of \texttt{astropy} \citep[][]{astropy:2022}, \texttt{numpy} \citep[][]{harris2020array}, \texttt{matplotlib} \citep[][]{Hunter:2007}, \texttt{emcee} \citep[][]{emcee_13} and \texttt{dynesty} \citep[][]{Speagle_20}. 
\end{acknowledgements}

\bibliographystyle{aa}
\bibliography{bib}

\end{document}